\documentclass[a4paper,11pt]{article}
\usepackage{jcappub} 
\usepackage{graphicx}
\usepackage{amsmath,amsthm,amssymb}
\usepackage{bm}
\usepackage{amsfonts}
\usepackage{url}

\def\be{\begin{equation}}
\def\ee{\end{equation}}
\def\ba{\begin{eqnarray}}
\def\ea{\end{eqnarray}}
\def\com{\color{magenta}}
\def\cob{\color{blue}}

\newcommand{\C}{{\cal C}}
\newcommand{\D}{{\cal D}}

\newcommand{\aH}{\alpha_{\rm H4}}

\newcommand{\at}{\alpha_{\rm t4}}
\newcommand{\arX}[1]{\href{http://arxiv.org/abs/#1}{{\ttfamily\com arXiv:#1}}}
\newcommand{\doij}[5]{\href{http://dx.doi.org/#1}{\cob {\it #2} {\bf #3} (#5) #4}}

\title{Conical singularities and the Vainshtein screening 
in full GLPV theories}

\author[a]{Ryotaro Kase,}
\author[a]{Shinji Tsujikawa}
\author[b]{and Antonio De Felice}

\affiliation[a]{Department of Physics, Faculty of Science,
Tokyo University of Science, 1-3,
Kagurazaka, Shinjuku, Tokyo 162-8601, Japan}
\affiliation[b]{Yukawa Institute for Theoretical Physics,
Kyoto University, 606-8502, Kyoto, Japan}

\emailAdd{r.kase@rs.tus.ac.jp}
\emailAdd{shinji@rs.kagu.tus.ac.jp}
\emailAdd{antonio.defelice@yukawa.kyoto-u.ac.jp}

\abstract{In Gleyzes-Langlois-Piazza-Vernizzi (GLPV) theories, 
it is known that the conical singularity arises 
at the center of a spherically symmetric body ($r=0$) 
in the case where the parameter $\alpha_{{\rm H}4}$ 
characterizing the deviation from the Horndeski Lagrangian 
$L_4$ approaches a non-zero constant as $r \to 0$. 
We derive spherically symmetric solutions around 
the center in full GLPV theories and show that the GLPV 
Lagrangian $L_5$ does not modify 
the divergent property 
of the Ricci scalar $R$ induced by the non-zero
$\alpha_{{\rm H}4}$. 
Provided that $\alpha_{{\rm H}4}=0$, 
curvature scalar quantities can remain finite at $r=0$ 
even in the presence of $L_5$ 
beyond the Horndeski domain.
For the theories in which the scalar field $\phi$ 
is directly coupled to $R$, we also obtain 
spherically symmetric solutions inside/outside 
the body to study whether the fifth force 
mediated by $\phi$ can be screened by non-linear 
field self-interactions. We find that there is one
specific model of GLPV theories in which the 
effect of $L_5$ vanishes in the equations of motion.
We also show that, depending on the sign of 
a $L_5$-dependent term in the field equation, 
the model can be compatible with solar-system 
constraints under the Vainshtein mechanism 
or it is plagued by the problem of a divergence of 
the field derivative in high-density regions.
}

\begin{document}
\maketitle
\flushbottom

\section{Introduction}
\label{intro} 

Over the past century, General Relativity (GR) has 
persistently been the fundamental theory for describing 
the gravitational interaction with matter. 
The standard model of particle physics has also been 
the backbone of high energy physics along 
with the discovery of the Higgs particle. 
Nevertheless, unifying gravity with other forces in Nature  
is still a challenging task. 
Moreover, the observational evidence of a
late-time  cosmic acceleration \cite{Riess,Perl}  
implies that the gravitational law may be modified on 
infra-red scales to realize an effective negative pressure against 
gravity \cite{Sil,Tsuji10,Pedro}.  

The cosmological constant arising from the vacuum 
energy of particle physics can lead to the cosmic acceleration, 
but its energy scale is vastly larger than the observed 
dark energy scale \cite{Weinberg}. 
An alternative approach to explaining 
the Universe acceleration (for both dark energy and inflation) 
is to introduce a scalar field $\phi$ beyond the standard model 
of particle physics \cite{CST}. Such a scalar field can be
generally coupled to gravity \cite{Brans,Fujiibook} 
through interactions like $G_4(\phi,X)R$ and 
$G_5(\phi,X)G_{\mu \nu}\nabla^{\mu}\nabla^{\nu}\phi$, 
where $G_{4,5}(\phi,X)$ are functions of $\phi$ and 
$X=\nabla_{\mu}\phi\nabla^{\mu}\phi$ ($\nabla_{\mu}$ 
is the covariant derivative operator), 
$R$ is the Ricci scalar, and 
$G_{\mu \nu}$ is the Einstein tensor.

The $X$-dependent couplings with $R$ and $G_{\mu \nu}$ 
can give rise to equations of motion containing derivatives 
higher than second order \cite{covariant}. 
In such cases, the ghost-like Ostrogradski
instability \cite{Ostro} arises, in general, 
due to the Hamiltonian being unbounded from below.  
One can formulate the action of the most general 
scalar-tensor theories with second-order equations of motion 
by adding terms that cancel higher-order derivatives \cite{Deffayet}.  
Such a general action was first derived by Horndeski 
in 1974 \cite{Horndeski} and it received much attention 
after 2011 in connection to the problems of inflation and 
dark energy \cite{KYY,Char,Gao,DT11,DKT,Kimura}.

In Horndeski theories, there is only one propagating scalar 
degree of freedom (DOF) besides the two graviton polarizations. 
If we go beyond the Horndeski domain, then the theories have
derivatives higher than second order, but this does not necessarily 
increase the number of propagating DOF. 
In fact, Gleyzes-Langlois-Piazza-Vernizzi (GLPV) expressed the
Horndeski Lagrangian in terms of the 3+1 Arnowitt-Deser-Misner (ADM)
decomposition of space-time \cite{ADM} by choosing the unitary gauge
\cite{building} and derived more general theories without imposing the 
two conditions that Horndeski theories satisfy \cite{GLPV}.  According to
the Hamiltonian analysis in the unitary gauge, the GLPV theories have
only one scalar DOF \cite{GLPV,Lin,GLPV2,Gao2}.  The recent full 
gauge-invariant analysis of Ref.~\cite{counting} also supports this claim.

The GLPV theories were applied to the spherically symmetric background
\cite{Gel,Koba,Sakstein,Mizuno,Jimenez,AKT15,Sakstein2} as well as to
the cosmological setup relevant to dark energy
\cite{KT14,Koyama,GLMV1,weak,Perenon,Tsuji15,GLMV2,KTD15}.  
The main motivation for the derivation of spherically symmetric solutions 
is to understand how the screening of the fifth force mediated by the scalar
DOF works to recover the General Relativistic behavior inside the
solar system \cite{Babichev}.  In Horndeski theories up to the
Lagrangian $L_4$ (involving the non-minimal coupling $G_4(\phi,X)R$),
it is known that non-linear scalar-field self-interactions associated
with Galileons \cite{Nicolis,covariant} can lead to the suppression of
the fifth force around a compact body
\cite{GSami,Burrage,Brax,FKT12,Koba} through the Vainshtein mechanism
\cite{Vainshtein}.  The Horndeski Lagrangian $L_5$, which contains the
derivative coupling
$G_5(\phi,X)G_{\mu \nu}\nabla^{\mu}\nabla^{\nu}\phi$, 
generally prevents the success of the Vainshtein mechanism in 
high-density regions \cite{Kimura,Niz,KT13}.

In GLPV theories, 
up to the Lagrangian $L_4$, it was found that a 
conical singularity arises at the center of a spherically symmetric
body ($r=0$) for the models in which the parameter $\aH$,
characterizing the departure from Horndeski theories, 
approaches a non-zero constant as $r \to 0$ \cite{AKT15}. 
This singularity is related with the geometric modification of the
four-dimensional Ricci scalar $R$ expressed in terms of the three
dimensional extrinsic and intrinsic curvatures appearing in the ADM
formalism.  To avoid the divergence of curvature invariants 
at $r=0$ for spherically symmetric solutions, 
the deviation parameter $\aH$ was required to vanish for $r \to 0$
(whereas $\aH$ can be non-zero for $r>0$).  In fact, it is possible to
construct dark energy models without the conical singularity in GLPV
theories, while satisfying cosmological and local gravity constraints
\cite{KTD15}.

Since the analysis of Ref.~\cite{AKT15} is restricted to the
up-to-$L_4$ sub-class of GLPV theories, it remains to be seen what
happens around the center of a compact body in the full GLPV theories
containing the Lagrangian $L_5$.  This is the issue addressed in the
present paper. 
In particular, our first interest is to show whether the conical
singularity induced by the non-vanishing $\aH$ at $r=0$ is affected by
the addition of $L_5$.  Secondly, we would like to see whether the
Lagrangian $L_5$ outside the Horndeski domain can induce other
curvature singularities at $r=0$. 
In order to avoid that the divergence of curvature invariants 
(such as $R$) may occur because of the unphysical choice of free functions on
spherically symmetric backgrounds, we consider those GLPV
theories whose action remains finite in the limit $X \to 0$ with the regular
boundary condition $d\phi/dr=0$ at $r=0$.

In addition to the derivation of spherically symmetric 
solutions around $r=0$, we also obtain solutions 
inside/outside the compact body by taking into account
a non-minimal coupling between $\phi$ and $R$. 
This is for the purpose of addressing 
whether the Vainshtein mechanism is at work 
in the presence of the GLPV Lagrangian $L_5$.
In the GLPV domain, there is one particular 
model in which the coefficient $s_5$ arising from 
$L_5$ disappears in the field equations of motion. 
Depending on the sign of $s_5$, we find that the 
model can be compatible with solar-system constraints 
under the operation of the Vainshtein mechanism 
or it suffers from divergent behavior of the field derivative 
in regions of high density. 

This paper is organized as follows.  In Sec.~\ref{EOM} the equations
of motion in full GLPV theories are derived on the spherically
symmetric background by introducing quantities associated with the
Lagrangian in the ADM formalism.  In Sec.~\ref{centersec} we first
review how the conical singularity arises in the theories up to the
Lagrangian $L_4$ and then obtain general conditions for avoiding the
divergence of curvature invariants at $r=0$.  In Sec.~\ref{L5sec} we
study how the existence of the Lagrangian $L_5$ in both Horndeski and
GLPV theories affects the problem of the conical singularity at $r=0$.
In Sec.~\ref{Vasec} we derive spherically symmetric solutions both
inside and outside the body on account of the non-minimal coupling
between $\phi$ and $R$. We discuss conditions under which the model
with the Lagrangian $L_5$ can be compatible with local gravity
constraints through the operation of the Vainshtein mechanism.
Sec.~\ref{consec} is devoted to conclusions.

\section{Background equations of motion}
\label{EOM} 

The action of GLPV theories \cite{GLPV} is given by 
\be
S=\int d^4 x \sqrt{-g} \sum^{5}_{i=2}L_i
+\int d^4 x \sqrt{-g}\,L_m (g_{\mu \nu},\Psi_m)\,,
\label{act}
\ee
where $g$ is the determinant of the metric $g_{\mu \nu}$, 
$L_m$ is a matter Lagrangian with 
matter fields $\Psi_m$, and 
\ba
L_{2} &=&G_{2}(\phi,X)\,,  
\label{L2} \\
L_{3} &=&G_{3}(\phi,X) \square \phi\,,  
\label{L3} \\
L_{4} &=&G_{4}(\phi,X) R-2G_{4,X}(\phi,X)
\left[ (\square \phi )^{2}-\nabla ^{\mu} \nabla^{\nu}\phi 
\nabla_{\mu}\nabla_{\nu}\phi \right]  \notag \\
& &+F_{4}(\phi,X)\epsilon^{\mu\nu\rho\sigma}\epsilon_{\mu'\nu'\rho'\sigma} 
\nabla^{\mu'}\phi\nabla_{\mu}\phi
\nabla^{\nu'}\nabla_{\nu}\phi
\nabla^{\rho'}\nabla_{\rho}\phi\,,  \label{L4}\\
L_{5} &=&G_{5}(\phi,X)G_{\mu\nu}\nabla^{\mu}\nabla^{\nu}\phi \notag \\
& &+\frac{1}{3}G_{5,X}(\phi,X)
\left[(\Box\phi)^{3}
-3\Box\phi\nabla_{\mu}\nabla_{\nu}\phi\nabla^{\mu}\nabla^{\nu}\phi
+2\nabla_{\mu}\nabla_{\nu}\phi\nabla^{\sigma}\nabla^{\mu}\phi
\nabla_{\sigma}\nabla^{\nu}\phi\right]\notag \\
&  & +F_{5}(\phi,X)\epsilon^{\mu\nu\rho\sigma}\epsilon_{\mu'\nu'\rho'\sigma'} 
\nabla^{\mu'}\phi\nabla_{\mu}
\phi\nabla^{\nu'}\nabla_{\nu}\phi
\nabla^{\rho'}\nabla_{\rho}\phi
\nabla^{\sigma'}\nabla_{\sigma}\phi
\,. \label{L5}
\ea
Here $G_i(\phi,X)$ $(i=2, 3, 4, 5)$ and $F_j(\phi,X)$ $(j=4, 5)$ are
functions of the scalar field $\phi$
and its kinetic term $X\equiv \nabla^{\mu}\phi\nabla_{\mu}\phi$, $R$
is the Ricci scalar, $G_{\mu\nu}$ is the Einstein tensor,
$\epsilon_{\mu\nu\rho\sigma}$ is the totally antisymmetric Levi-Civita
tensor, and a comma represents the partial derivative with respect to
$\phi$ or $X$, e.g., $G_{4,X}\equiv\partial G_4/\partial X$.  We use
the metric signature $(-,+,+,+)$.  Horndeski theories correspond to
the choice $F_4=F_5=0$. 

The action (\ref{act}) was discovered by expressing the Horndeski
action in terms of the 3+1 ADM decomposition of space-time by choosing
the unitary gauge ($\phi=\phi(t)$) for a time-like scalar field
\cite{building,GT14,Tsuji2014lec} and by generalizing it without 
imposing the two conditions Horndeski theories obey \cite{GLPV}. 
In terms of the extrinsic curvature $K_{\mu\nu}$ and the intrinsic
curvature ${\cal R}_{\mu\nu}$, the Lagrangian of GLPV theories 
on such a background is simply given by
\be
L=A_2+A_3K+A_4(K^2-K_{\mu\nu}K^{\mu \nu})
+B_4 {\cal R}+A_5{\cal K}
+B_5 \left( K^{\mu \nu}{\cal R}_{\mu\nu}
-\frac{K{\cal R}}{2} \right)\,,
\ee
where $K\equiv g^{\mu\nu}K_{\mu\nu}$,
${\cal K}\equiv K^3-3KK^{\mu\nu}K_{\mu\nu}
+2K_{\mu\nu}K^{\mu\rho}{K^{\nu}}_{\rho}$,
and ${\cal R}\equiv g^{\mu\nu}{\cal R}_{\mu\nu}$.  
The functions $A_{2,3,4,5}$ and $B_{4,5}$ are related 
with $G_{2,3,4,5}$ and $F_{4,5}$ in 
Eqs.~(\ref{L2})-(\ref{L5}) according to
\ba
A_2&=&G_2-XE_{3,\phi}\,,\label{A2}\\
A_3&=&2|X|^{3/2}\left(E_{3,X}+\frac{G_{4,\phi}}{X}\right)\,,\label{A3}\\
A_4&=&-G_4+2XG_{4,X}+\frac{X}{2}G_{5,\phi}-X^2F_{4}\,,\label{A4}\\
B_4&=&G_4+\frac{X}{2}(G_{5,\phi}
-E_{5,\phi})\,,\label{B4}\\
A_5&=&-|X|^{3/2}\left(XF_5+\frac{G_{5,X}}{3}\right)\,,\label{A5}\\
B_5&=&\frac{|X|^{3/2}}{X}E_5\,,\label{B5}
\ea
where $E_{3}(\phi,X)$ and $E_{5}(\phi,X)$ are 
auxiliary functions satisfying 
\be
G_3=E_3+2XE_{3,X}\,,\qquad 
G_{5,X}=\frac{E_5}{2X}+E_{5,X}\,.
\label{E5def}
\ee
These auxiliary functions were introduced in Ref.~\cite{building} for
the convenience of rewriting Horndeski Lagrangians in the ADM form.
The relations (\ref{A2})-(\ref{B5}) were first derived for the
isotropic cosmological background with $X<0$ \cite{GLPV}.  Even for
the background with $X>0$, we can use the functions $A_{2,3,4,5}$ and
$B_{4,5}$ in (\ref{A2})-(\ref{B5}) to obtain equations of motion
simpler than those derived with the functions $G_{2,3,4,5}$ and
$F_{4,5}$. Since our interest in this paper is the spherically
symmetric solutions, we focus on the case $X>0$ in the following
discussion.

We have the following relations 
\ba
& &
A_4+B_4-2XB_{4,X}=-X^2F_4\,,\\
& &
A_5+\frac13 XB_{5,X}=-X|X|^{3/2}F_5\,.
\ea
The two relations, $F_4=0$ and $F_5=0$, which define the Horndeski theories,
correspond to $A_4+B_4-2XB_{4,X}=0$ and $A_5+XB_{5,X}/3=0$,
respectively.  Therefore, the functions $F_4$ and $F_5$ characterize the
deviation from Horndeski theories. 

We consider the spherically symmetric and static 
space-time described by the line-element 
\be
ds^{2}=-e^{2\Psi(r)}dt^{2}+e^{2\Phi(r)}dr^{2}
+r^{2} \left( d\theta^{2}+\sin^{2}\theta\, 
d\varphi^{2} \right)\,,
\label{line}
\ee
where $\Psi(r)$ and $\Phi(r)$ are the gravitational potentials 
depending on the radius $r$ from the center of symmetry. 
As for the matter Lagrangian we consider a perfect fluid whose 
energy-momentum tensor is given by 
$T^{\mu}_{\nu}={\rm diag}\left(-\rho_m,P_m,P_m,P_m \right)$, 
where $\rho_m$ is the energy density and $P_m$ is the pressure.
Variation of the action (\ref{act}) leads to the following 
equations of motion
\ba
&&\left( \frac{4 e^{-2 \Phi} A_4}{r}
-\C_1+4\C_2+\D_1 +\D_2\right) \Phi'
-A_2+\C_3-\C_4-\D_3\notag\\
&&-\frac{2 (e^{-2\Phi}-1-\at)A_4}{r^2}
=-\rho_m
\,,\label{e00}\\
&&\left( \frac{4 e^{-2 \Phi} A_4}{r}
-\C_1+4\C_2+\D_1 +\D_2 \right) \Psi' 
+A_2-2XA_{2,X}-\frac{2(\C_1-\C_2)}{r}\notag\\
&&+\frac{2 (e^{-2\Phi}-1-\aH)A_4}{r^2}
=-P_m\,,\label{e11}\\
&&\left( 2e^{-2\Phi} A_4+{\cal D}_4 \right)(\Psi''+\Psi'^2) 
-\left[2 e^{-2\Phi}A_4+2\D_4
+r(2\C_2+\D_1)\right]\Psi'\Phi'
+A_2-\C_3+\frac{\C_4}{2}
\notag\\
&&+\left(\frac{2e^{-2\Phi}A_4}{r}+\frac{\C_4}{2}r
+\D_5\right)\Psi' 
-\left(\frac{2e^{-2\Phi}A_4}{r}-\C_1+2\C_2\right)\Phi'
=-P_m\,,
\label{e22}\\
&&P_m'+\Psi'(\rho_m+P_m)=0\,,\label{con}
\ea
where a prime represents a derivative 
with respect to $r$, and 
\ba
&&
\C_1\equiv2Xe^{-\Phi}A_{3,X}\,,\qquad 
\C_2\equiv\frac{2e^{-2\Phi}XA_{4,X}}{r}\,,\nonumber \\
& &\C_3\equiv e^{-\Phi}\phi'(A_{3,\phi}+2e^{-2\Phi}\phi'' A_{3,X})\,,\qquad 
\C_4\equiv\frac{4e^{-2\Phi}\phi'(A_{4,\phi}+2e^{-2\Phi}\phi'' A_{4,X})}{r}\,,\nonumber \\
& &\D_1=\frac{12e^{-3\Phi}XA_{5,X}}{r^2}\,,\qquad
\D_2=\frac{2e^{-\Phi}}{r^2}
\left( 6e^{-2\Phi}A_5+XB_{5,X} \right)\,,\nonumber \\
& &
\D_3\equiv\frac{2e^{-3\Phi}\phi'\phi'' (6e^{-2\Phi}A_{5,X}+B_{5,X})
+e^{-\Phi}\phi'(6e^{-2\Phi}A_{5,\phi}+B_{5,\phi})}
{r^2}\,,\nonumber \\ 
& &\D_4=\frac{6e^{-3\Phi}A_5}{r}\,,\qquad
\D_5\equiv\frac{6e^{-3\Phi}\phi'(A_{5,\phi}
+2e^{-2\Phi}\phi'' A_{5,X})}{r}\,,
\label{coeff}
\ea
and 
\ba
\at &\equiv&-\frac{B_4}{A_4}-1\,,\\
\aH &\equiv& \frac{2XB_{4,X}-B_4}{A_4}-1
=\frac{X^2F_4}{A_4}\,.
\label{a4def}
\ea
The parameter $\aH$ characterizes the deviation of the GLPV theories
from the Horndeski domain for the Lagrangians up to $L_4$.  
In the presence of $L_5$, the difference between $A_5$ and 
$-XB_{5,X}/3$ also gives rise to the departure from 
the Horndeski domain, but we will
show that the deviation from the $L_4$ sector is crucial for the 
appearance of the conical singularity.

\section{Appearance of the conical singularity 
at the center of a compact body}
\label{centersec} 

The conical singularity found in Ref.~\cite{AKT15} appears at the
center of a compact body in the case where the parameter $\aH$
approaches a non-vanishing constant as $r \to 0$.  
This happens for the simple model
in which $-A_4$ and $B_4$ are non-zero constants different from each
other.  In the following, we first review how the conical singularity
arises in such a simple model and then discuss general conditions for
avoiding the divergence of curvature scalar quantities.

\subsection{Models with a conical singularity}
\label{L4sec}

Let us begin with the model of a massless scalar field characterized
by the functions
\be
A_2=-\frac12 X\,,\quad A_3=0\,,\quad
A_4=-\frac{1}{1+\aH}G_4\,,\quad B_4=G_4\,,
\quad A_5=B_5=0\,,
\label{model1}
\ee
where $G_4$ and $\alpha_{{\rm H}4}~(= \alpha_{{\rm t}4})$ are
constants. This is arguably the simplest extension of GR 
($-A_4=B_4=M_{\rm pl}^2/2$, where $M_{\rm pl}$ is the reduced Planck
mass) to the domain of GLPV theories ($\aH \neq 0$).

We consider a spherically symmetric body and derive 
the solutions to Eqs.~(\ref{e00})-(\ref{con}) 
around its center ($r=0$). 
In doing so, we expand the gravitational potentials 
and the scalar field around $r=0$, as 
\ba
\Phi(r) &=& \Phi_0+\sum_{i=1}^{\infty} \Phi_i r^i\,, 
\label{Phir}\\
\Psi(r) &=& \Psi_0+\sum_{i=1}^{\infty} \Psi_i r^i\,, 
\label{Psir}\\
\phi(r) &=& \phi_0+\sum_{i=2}^{\infty} \phi_i r^i\,,
\label{phir}
\ea
where $\Phi_0$, $\Phi_i$, $\Psi_0$, $\Psi_i$, $\phi_0$, and $\phi_i$
are constants.  In Eq.~(\ref{phir}) the linear term proportional to
$r$ is absent to respect the regular boundary condition $\phi'(0)=0$.
In GR the linear terms of $\Phi(r)$ and $\Psi(r)$ also vanish. This
will in general happen also for the GLPV theories, as we shall see
later on.

For constant $\rho_m$, we can integrate Eq.~(\ref{con}) to give
\be
P_m(r)=-\rho_m+\rho_s e^{-\Psi(r)}\,,
\label{Pmso}
\ee
where $\rho_s$ is a constant.  If the density depends on $r$, we can
expand $\rho_m(r)$ in the form $\rho_m(r)=\rho_0+\rho_1r^2+\cdots$
around $r=0$. This difference does not affect the discussion of the conical
singularity given below, so we shall focus on the case of constant
$\rho_m$ up to the end of Sec.~\ref{L5sec}.  In Sec.~\ref{Vasec} we
study the case in which the density varies as a function of $r$ in
order to discuss the behavior of the solutions inside and outside 
the body. 

Substituting Eqs.~(\ref{Phir})-(\ref{Pmso}) into
Eqs.~(\ref{e00})-(\ref{e22}), we can iteratively derive the
coefficients of $\Phi(r)$, $\Psi(r)$, and $\phi(r)$.  For the model
(\ref{model1}) we obtain the following solutions
\ba
\Phi(r) &=& -\frac12 \ln \left( 1+\aH \right) 
+\frac{\rho_m}{12G_4}r^2+
\frac{\rho_m^2}{144G_4^2}r^4
+{\cal O}(r^6)\,,
\label{Phiso} \\
\Psi(r) &=& \Psi_0-\frac{2\rho_m e^{\Psi_0}-3\rho_s}
{24G_4 e^{\Psi_0}}r^2
-\frac{(4e^{\Psi_0}\rho_m-3\rho_s)
(2e^{\Psi_0}\rho_m-3\rho_s)}
{1152G_4^2e^{2\Psi_0}}r^4
+{\cal O}(r^6)\,,
\label{Psiso}\\
\phi(r)&=&\phi_0\,.
\label{phiso}
\ea
Since the coefficients $\phi_i$ (where $i \geq 2$) vanish, 
the scalar field is constant around $r=0$. 
On using these solutions, the Ricci scalar is given by 
\be
R=-\frac{2\aH}{r^2}
+\frac{(4\rho_m e^{\Psi_0}-3\rho_s)(1+\aH)}
{2G_4e^{\Psi_0}}+{\cal O}(r^2)\,.
\ee
For the theories with $\aH \neq 0$, there is the conical singularity
associated with the divergence of $R$.  This singularity arises
independently of the choice of the integration constants $\rho_s$,
$\Psi_0$, and $\phi_0$.

\subsection{Requirements for the absence of 
singularities at the origin}
\label{resec}

In Sec.~\ref{L4sec} we have shown the existence of the conical
singularity for the simple model (\ref{model1}), but this property is
generic for the models in which the parameter $\aH$ approaches a
non-zero constant as $r \to 0$.  On using the expansions
(\ref{Phir})-(\ref{phir}) for general models, the Ricci scalar $R$
around $r=0$ reads
\be
R = 
\frac{2(1-e^{-2\Phi_0})}{r^2}
+\frac{4(2\Phi_1-\Psi_1)e^{-2\Phi_0}}{r}
-2(6\Phi_1^2+\Psi_1^2-5\Phi_1 \Psi_1
-6\Phi_2+6\Psi_2)e^{-2\Phi_0}+{\cal O}(r)\,. 
\label{Rso}
\ee
Thus, the gravitational potentials up to first order in $r$ are
intrinsically related to the divergence of $R$.  The Ricci scalar
remains finite under the two conditions
\ba
\Phi_0 &=& 0\,,\label{con1d} \\
2\Phi_1 &=&\Psi_1\,. \label{con2d}
\ea

On account of the two conditions (\ref{con1d}) and (\ref{con2d}), we
also find
\ba
R_{\mu\nu}R^{\mu\nu}
&=&
{\frac {{22\Phi_{{1}}^2}}{{r}^{2}}}-{\frac {8\Phi_{{1}} 
\left( 9\,{\Phi_{{1}}^2}-4\,\Phi_{{2}}-4\,
\Psi_{{2}} \right) }{r}}
+\frac{400}{3}\Phi_1^4-8\Phi_1^2 
\left( 25\Phi_2+11\Psi_2 \right) \nonumber \\
& &
+48\left( \Phi_2^2 -\Phi_2\Psi_2+\Psi_2^2 \right)
+4\Phi_1 \left( 11\Phi_3+15\Psi_3 \right)+{\cal O}(r)\,, \\
R_{\mu\nu\rho\sigma}R^{\mu\nu\rho\sigma}
&=&
{\frac {{56\Phi_{{1}}^2}}{{r}^{2}}}-{\frac {64\Phi_{{1}} 
\left( 3\,{\Phi_{{1}}^2}-\Phi_{{2}}-\Psi_{{2
}} \right) }{r}}+\frac{1120}{3}\Phi_1^4
-32\Phi_1^2 \left( 12\Phi_2+7\Psi_2 \right) 
\nonumber \\
& &+48\left( \Phi_2^2+\Psi_2^2 \right)
+16\Phi_1 \left( 5\Phi_3+6\Psi_3 \right)+{\cal O}(r)\,,
\label{Ricciso}
\ea
so that the singularity is absent at $r=0$ if and only if
$\Phi_1=0=\Psi_1$, as it also happens in GR. 

In summary, the static object does not possess curvature 
singularities at $r=0$ if 
\ba
\Phi_0 &=& 0\,,\label{con1} \\
\Phi_1 &=& 0\,, \label{con2}\\
\Psi_1 &=& 0\,. \label{con3}
\ea
For the solutions (\ref{Phiso})-(\ref{Psiso}) we have that
$e^{-2\Phi_0}=1+\aH$ and $\Phi_1=\Psi_1=0$, 
so the first condition (\ref{con1}) is violated for $\aH \neq 0$.

In GLPV theories up to the Lagrangians $L_4$, one can generalize the
model (\ref{model1}) in such a way that additional contributions to
the functions $A_2,A_3,A_4, B_4$ are present.  
Introducing the cosmological constant $-\Lambda$ 
in $A_2$, we have additional $r^2$-dependent contributions 
to Eqs.~(\ref{Phiso})-(\ref{Psiso}). However, 
this does not modify the existence of the conical singularity. 

If we take into account additional $X$-dependent terms like 
$A_i \propto X^i$ (where $i \geq 1$), it follows that the coefficients 
${\cal C}_{1,2,3}$ in Eq.~(\ref{coeff}) vanish for $r \to 0$ 
with the field profile (\ref{phir}) satisfying 
the boundary condition $\phi'(0)=0$. On using the properties 
$\phi'(r) \simeq 2\phi_2r$ and $\phi''(r) \simeq 2\phi_2$ 
around $r=0$ for the function $A_4$ containing $\beta_4 X$ 
($\beta_4$ is a constant), the coefficient ${\cal C}_4$ 
in Eq.~(\ref{coeff}) approaches a constant value 
$32\beta_4\phi_2^2e^{-4\Phi_0}$ as $r \to 0$. 
Apparently the constant ${\cal C}_4$ can be absorbed into 
$A_2$ in Eq.~(\ref{e00}), but we need to caution that 
it appears differently in Eq.~(\ref{e22}) 
with an extra contribution ${\cal C}_4/2$ relative to $A_2$.
To address this issue, we first derived the solutions to 
$\Phi$ and $\Psi$ iteratively by expanding 
Eqs.~(\ref{e00})-(\ref{e11}) up to quadratic order 
in $r$ and then substituted them into Eq.~(\ref{e22}). 
This second step leads to the equality 
${\cal C}_4/2+c\phi_2^2r^2+{\cal O}(r^3)=0$, where 
$c$ is a non-zero constant. 
Since the term $c\phi_2^2r^2$ needs to vanish, 
we obtain $\phi_2=0$. 
Under this condition the first term 
${\cal C}_4/2=16\beta_4\phi_2^2e^{-4\Phi_0}$ 
vanishes identically, so there is no inconsistency 
for the extra term ${\cal C}_4/2$ mentioned above.
For the theories with additional terms $-\Lambda$ in $A_2$ and 
$\beta_4 X$ in $A_4$ to the functions (\ref{model1}),  
the resulting gravitational 
potentials correspond to Eqs.~(\ref{Phiso})-(\ref{Psiso}) 
with the replacement $\rho_m \to \rho_m+\Lambda$. 

Another possibility is to take into account the $\phi$-dependence in
the functions $A_2,A_3,A_4$, and $B_4$.  Since $\phi(r)$ approaches a
constant as $r \to 0$, an additional $\phi$-dependent contribution to
$A_2$ behaves as the cosmological constant $\Lambda$ discussed above.
Similarly, in the limit that $r \to 0$, ${\cal C}_3$ vanishes and
${\cal C}_4$ approaches a constant, so addition of $\phi$-dependent
terms in $A_3$ and $A_4$ does not affect the existence of the conical
singularity.  We note that the $\phi$-dependence in $A_4$ and $B_4$
leads to the variation of the scalar field such that $\phi_2$ is a
non-zero constant in Eq.~(\ref{phir}) \cite{AKT15} (as we will see in
Sec.~\ref{Vasec}).  However, this does not modify the solutions
(\ref{Phiso})-(\ref{Psiso}) up to first order in $r$.

\section{Effect of the Lagrangian $L_5$ 
on the conical singularity}
\label{L5sec} 

We study how the Lagrangian $L_5$ affects the problem of the conical
singularity discussed in Sec.~\ref{centersec}.  The quantities $A_5$
and $B_5$ appearing in Eqs.~(\ref{e00})-(\ref{e22}) are related with
the functions $G_5$ and $F_5$ in the Lagrangian (\ref{L5}), as
Eqs.~(\ref{A5}) and (\ref{B5}).  Let us consider the functional form
of $G_5(\phi,X)$ given by
\be
G_5(\phi,X)=\sum_{n=0} g_{n/2}(\phi)X^{n/2}\,,
\label{G5ex}
\ee
where $g_{n/2}(\phi)$ is an arbitrary function of $\phi$, and
$n~(\geq 0)$ is an integer.  The choice of the power $n/2$ in
$X=e^{-2\Phi}{\phi}'^{2}$ comes from the fact that we would like to
take into account the terms with odd powers of ${\phi}'$ in addition
to even powers.  The function (\ref{G5ex}) can encompass most of the
models with the Lagrangian $L_5$ proposed in the literature.  For
example, we have $G_5(\phi)=g_0(\phi) \propto \phi$ for the
non-minimal derivative coupling model \cite{Luca,Germani},
$G_5(X)=g_2 X^2$ for covariant Galileons \cite{covariant}, and
$G_5(X)=g_m X^m$ ($m \geq 1$) for extended Galileons \cite{extended}.

For the purpose of studying the singularity problem around the center
of a compact body, the function (\ref{G5ex}) is sufficiently general
as it can be regarded as the expansion of a regular function $G_5$
around $r=0$.  Note that, for the finiteness of the action (\ref{act})
in the limit $\phi' \to 0$, $G_5$ should not contain negative 
powers of $X^m$.  Since the leading-order term of the field derivative
is given by $\phi'(r) \propto r$, it follows that the r.h.s.\ of
Eq.~(\ref{G5ex}) can be expanded as $G_5=g_0+\sum_{i=1}G_{5i}\,r^i$,
where $G_{5i}$'s are constants.  This is of the same form as the
expansions (\ref{Phir})-(\ref{Psir}) of $\Phi(r)$ and $\Psi(r)$ around
$r=0$.

The auxiliary function $E_5$ defined by 
Eq.~(\ref{E5def}) reads
\be
E_5={\cal B}(\phi) X^{-1/2}
+\sum_{n=1} \frac{n}{n+1} 
g_{n/2} (\phi) X^{n/2}\,,
\ee
where ${\cal B}(\phi)$ is a function of $\phi$ arising from the
integral with respect to $X$. Since the arbitrariness of the function
${\cal B}(\phi)$ should not affect the background equations of motion
through the term $B_{5,\phi}$, we focus on the case in which
${\cal B}$ does not depend on $\phi$.  {}From Eqs.~(\ref{A5}) and
(\ref{B5}), it follows that
\ba
A_5 &=& -X^{5/2} F_5-\sum_{n=1}
\frac{n}{6}g_{n/2}(\phi) X^{(n+1)/2}\,,\label{A5ex}\\
B_5 &=& {\cal B}+\sum_{n=1}
\frac{n}{n+1}g_{n/2}(\phi) X^{(n+1)/2}\,.
\label{B5ex}
\ea
Since $F_5=0$ in Horndeski theories, the function $A_5$ does not
contain a non-zero constant term.  In GLPV theories it is possible to
have constant $A_5$ for the choice $F_5(\phi)=c_5 X^{-5/2}$, where
$c_5$ is a constant.  For the finiteness of the action (\ref{act}) in
the limit $X \to 0$, we require that both $A_5$ and $B_5$ do not
involve the negative powers of $X^m$ [which is the case for $B_5$ in
Eq.~(\ref{B5ex})].  As we will see later in Sec.~\ref{nonF5}, this
restricts the functional form of $F_5$.

The existence of the term $G_5(\phi,X)$ also gives rise to
modifications to the functions $A_4$ and $B_4$, as
\ba
A_4 &=& \frac{1}{1+\aH}
\left[ -G_4+2XG_{4,X}
+ \sum_{n=0}\frac12g_{n/2,\phi}(\phi) 
X^{n/2+1} \right]\,,\label{A4con} \\
B_4 &=& G_4
+\sum_{n=0} \frac{1}{2(n+1)}
g_{n/2,\phi}(\phi) X^{n/2+1}\,,
\label{B4con}
\ea
where we used Eq.~(\ref{a4def}).  For example, the non-minimal
derivative coupling model corresponds to $g_0(\phi)=c_0\phi$ and
$g_{n/2}(\phi)=0$ for $n \geq 1$.  In this case, the terms
proportional to $X$ are present in Eqs.~(\ref{A4con}) and
(\ref{B4con}).  As we already discussed in Sec.~\ref{resec}, this
additional term does not modify the existence of the conical
singularity at $r=0$.  This is also the case for the function
(\ref{G5ex}) containing the terms with the power $n \geq 1$.

\subsection{$F_5=0$}

To study the effect of the terms $A_5$ and $B_5$ on the equations of
motion (\ref{e00})-(\ref{e22}), we first discuss the case
$F_5=0$. Picking up one term in Eq.~(\ref{G5ex}), i.e.,
$G_5(\phi,X)=g_{n/2}(\phi)X^{n/2}$, it follows that
\ba
A_5 &=& -\frac{n}{6}g_{n/2}(\phi) X^{(n+1)/2}\,,
\label{A50}\\
B_5 &=& {\cal B}+\frac{n}{n+1}g_{n/2}(\phi) 
X^{(n+1)/2}\,,
\label{B50}
\ea
where $n \geq 1$.  For $n=0$ the terms $A_5$ and $B_5$ do not provide
any contribution to the equations of motion, but the effect of $L_5$
appears through the functions $A_4$ and $B_4$. As discussed in
Sec.~\ref{L4sec}, we consider the model of a massless scalar field
characterized by the functions
\ba
&& A_2=-\frac12 X\,,\qquad A_3=0\,,\nonumber \\
& &A_4=\frac{1}{1+\aH} \left[ -G_4
+\frac12 g_{n/2,\phi}(\phi) X^{n/2+1} \right]\,,\nonumber \\
& &B_4=G_4+\frac{1}{2(n+1)}
g_{n/2,\phi}(\phi) X^{n/2+1}\,,
\label{B40}
\ea
where $G_4~(\,\neq 0)$ and $\aH$ are constants.  The deviation from
Horndeski theories ($\aH \neq 0$) arises from the Lagrangian $L_4$.
For the theories with $n>2$, the term $X^{(n+1)/2}$ in
Eqs.~(\ref{A50}) and (\ref{B50}) has the power-law index larger than
$3/2$.  On using the fact that the field $\phi(r)$ can be expanded as
Eq.~(\ref{phir}) around $r=0$, the five coefficients
${\cal D}_{1,2,3,4,5}$ for $n>2$ vanish as $r \to 0$. Hence the
conical singularity induced by the non-zero constant $\aH$ term is not
affected by the terms with $n>2$ in $G_5$.  This includes the case of
the Lagrangian $L_5$ for covariant Galileons ($n=4$).

For the power $n=2$, it follows that the coefficients
${\cal D}_{1,2,4,5}$ vanish as $r \to 0$, whereas ${\cal D}_3$ behaves
as
\be
{\cal D}_3 \to 16 e^{-4\Phi_0} \left( 1-3 e^{-2\Phi_0} 
\right) \phi_2^3\,g_1(\phi_0)\,.
\ee
This term simply works as an additional constant to $A_2$ in
Eq.~(\ref{e00}), so the conical singularity is not affected by the
term $g_1(\phi)X$ in Eq.~(\ref{G5ex}).

For $n=1$, the coefficient ${\cal D}_4$ vanishes as $r \to 0$, while
the other coefficients behave as
\ba
& &
{\cal D}_1 \to -8 e^{-5\Phi_0}\phi_2^2\,g_{1/2}(\phi_0)\,,
\qquad 
{\cal D}_2 \to 4e^{-3\Phi_0}\left( 1-2 e^{-2\Phi_0} 
\right) \phi_2^2\,g_{1/2}(\phi_0)\,, \nonumber \\
& &
{\cal D}_3 \to \frac{4}{r}e^{-3\Phi_0}\left( 1-2 e^{-2\Phi_0} 
\right) \phi_2^2\,g_{1/2}(\phi_0)\,, \qquad
{\cal D}_5 \to -8 e^{-5\Phi_0}\phi_2^2\,g_{1/2}(\phi_0)\,.
\label{Dso}
\ea
In this case, we expand $\Phi, \Psi, \phi$ as
Eqs.~(\ref{Phir})-(\ref{phir}) and substitute the functions
(\ref{A50})-(\ref{B40}) and their derivatives into
Eqs.~(\ref{e00})-(\ref{e22}) to derive the solutions around $r=0$
iteratively.  In doing so, we consider the terms like
$g_{n/2}(\phi)$, $g_{n/2,\phi}(\phi)$, and $g_{n/2,\phi \phi}(\phi)$
as constants. 
Then, we obtain the same solutions as those derived in
Eqs.~(\ref{Phiso})-(\ref{phiso}).  Since the field $\phi(r)$ stays
constant, we have $\phi_2=0$ and hence all the coefficients of
Eq.~(\ref{Dso}) vanish.  Note that the solutions for $n \geq 2$ are
also given by Eqs.~(\ref{Phiso})-(\ref{phiso}).

The above discussion shows that, for $F_5=0$, the conical singularity
arising from the non-zero $\aH$ is not affected by the Lagrangian
$L_5$ containing the function $G_5$ in the form (\ref{G5ex}).

\subsection{$F_5 \neq 0$}
\label{nonF5}

Let us proceed to the case in which $F_5$ does not vanish (i.e.,
outside the Horndeski domain) with $G_5$ given by Eq.~(\ref{G5ex}).
Then, the functions $A_5,B_5,A_4,B_4$ are of the forms 
(\ref{A5ex})-(\ref{B4con}), where $G_4~(\,\neq 0)$ and $\aH$ are
assumed to be constant.  We also consider the massless scalar field
with $A_2=-X/2$ and $A_3=0$.

As long as $A_5$ does not contain negative powers 
of $X^m$, the action (\ref{act}) remains finite.
In addition to the positive powers originating from the 
sum on the r.h.s.\ of Eq.~(\ref{A5ex}), 
we can also take into account the terms $a_0(\phi)$ 
and $a_1(\phi)X^{1/2}$ from the new contribution 
$-X^{5/2}F_5$, where $a_0(\phi)$ and $a_1(\phi)$ 
are functions of $\phi$. 
In this case, we have 
\be
A_5=a_0(\phi)+a_1(\phi)X^{1/2}-\sum_{n=1}
\frac{n}{6}g_{n/2}(\phi) X^{(n+1)/2}\,.
\label{A5sum}
\ee
This expression can be regarded as the expansion 
of the regular function $A_5$ with respect to $r$ 
around the center of the compact body.

Now, we derive the solutions to Eqs.~(\ref{e00})-(\ref{e22}) under the
expansions (\ref{Phir})-(\ref{phir}) around $r=0$. After
differentiating the functions (\ref{A5ex})-(\ref{B4con}) with respect
to $\phi$, we deal with the terms such as $a_{0,\phi}$ and
$a_{1,\phi}$ as constants.  Up to linear order in $r$, the
gravitational potentials are given by
\ba
\Phi (r) &=&
\Phi_0+\frac16 \frac{(1+\aH-e^{-2\Phi_0})
e^{3\Phi_0} G_4}{a_0(1+\aH)}r+{\cal O}(r^2)\,,
\label{PhiGL}\\
\Psi (r) &=&
\Psi_0-\frac16 \frac{(1+\aH-e^{-2\Phi_0})
e^{3\Phi_0} G_4}{a_0(1+\aH)}r+{\cal O}(r^2)\,,
\label{PsiGL}
\ea
and $\phi_2=0$.  These solutions arise from the existence 
of the non-vanishing term $a_0$ in the limit $X \to 0$.  
The terms containing $\rho_m$ appear at cubic order in $r$. 
We recall that the curvature quantities are finite 
under the three conditions (\ref{con1}), (\ref{con2}), 
and (\ref{con3}).  Demanding the first
condition ($\Phi_0=0$) as the usual boundary condition in GR, the
coefficients of the linear term in $r$ read\footnote{It also follows
  that $\phi_i=0$ ($i \geq 3$), so the field $\phi(r)$ is constant.}
\be
\Phi_1=-\Psi_1=\frac{\aH}{6a_0(1+\aH)}G_4\,.
\ee
For the theories with $\aH \neq 0$, neither Eq.~(\ref{con2}) nor
Eq.~(\ref{con3}) are satisfied, so the divergence of curvature
quantities still persists even in the presence of the $F_5$ term.  If
$\aH=0$, then we have $\Phi_1=\Psi_1=0$ and hence there is no
divergent behavior of curvature invariants. 

Equations (\ref{PhiGL}) and (\ref{PsiGL}) look divergent in the limit that $a_0 \to 0$, but for the theories with $a_0=0$ the solutions to gravitational potentials are different from those 
for $a_0 \neq 0$.
When $a_0=0$ the dominant contributions to the coefficients 
of $\Phi'$ and $\Psi'$ in Eqs.~(\ref{e00})-(\ref{e11}) correspond to the term $4e^{-2\Phi}A_4/r$ around the origin. 
On the other hand, when $a_0 \neq 0$, the term 
$12e^{-3\Phi}A_5/r^2$, which comes from ${\cal D}_2$, dominates over the contribution 
$4e^{-2\Phi}A_4/r$ in the limit $r \to 0$.
In the latter case, the $1/r^2$ dependent term mentioned 
above gives rise to the unusual solutions (\ref{PhiGL}) 
and (\ref{PsiGL}) with an arbitrary value of $\Phi_0$. 
In other words, there is no continuous limit between 
the solutions for $a_0=0$ and $a_0 \neq 0$.
When $a_0=0$, the leading-order contribution to $\Phi$ 
is given by $\Phi_0= -(1/2)\ln (1+\alpha_{{\rm H}_4})$ without the divergences of $\Phi$ and $\Psi$. 
Hence, for $\alpha_{{\rm H}_4} \neq 0$, the conical singularity is present in this case as well.

The above discussion shows that the additional term $F_5$ to the
Lagrangian $L_5$ provides the new contributions $a_0(\phi)$ and
$a_1(\phi)X^{1/2}$ to $A_5$, but such terms themselves do not generate
new types of singularities at the origin.  The conical singularity
intrinsically arises for the models in which the parameter $\aH$
approaches a non-zero constant as $r \to 0$.  The existence of the
function $F_5$ beyond the Horndeski domain does not help to eliminate
the curvature singularity induced by the Lagrangian $L_4$ with
$\aH \neq 0$.  Since the functional forms (\ref{A5sum}) and
(\ref{B5ex}) can be regarded as the expansions of regular functions
$A_5$ and $B_5$ with positive powers of $X^m$, our results are valid
for general models in which the action (\ref{act}) is finite for
$X \to 0$.
 
\section{Solutions inside/outside the compact 
body and the screening mechanism}
\label{Vasec} 

In this section we obtain the field profile around the spherical
symmetric body under the approximation of weak gravity
($|\Phi| \ll 1$, $|\Psi| \ll 1$).  We take into account the
non-minimal coupling $G_4(\phi)R$ to study whether the screening of
the fifth force between the field $\phi$ and matter can be at work in
the presence of the Lagrangian $L_5$ with $F_5 \neq 0$. We also study
the matching of solutions inside and outside the body to place
constraints on the magnitude of the Lagrangian $L_5$ for the
consistency with local gravity experiments.

In GR ($-A_4=B_4=M_{\rm pl}^2/2$), the terms $4e^{-2\Phi}A_4\Phi'/r$
and $4e^{-2\Phi}A_4\Psi'/r$, which are of the order of $4A_4\Phi/r^2$,
are the dominant contributions to Eqs.~(\ref{e00}) and (\ref{e11}),
respectively.  We divide Eqs.~(\ref{e00})-(\ref{e11}) with respect to
$4A_4/r^2$ and consider each of the terms 
as small parameters at most of the order $\Phi$. 
This is required to recover the behavior close
to GR in the solar system \cite{FKT12,KT13}.  We also assume that the
parameters $|\aH|$ and $|\alpha_{{\rm t}4}|$ are much smaller than
unity and that the pressure $P_m$ of non-relativistic matter 
is of the same order as $\rho_m \Psi$.

Taking the $r$ derivative of Eq.~(\ref{e11}) and combining it with
Eq.~(\ref{con}), we can eliminate the $P_m'$ term to obtain the
second-order equation for $\Psi$.  Eliminating the second derivative
$\Psi''$ by using Eq.~(\ref{e22}) with Eqs.~(\ref{e00}) and
(\ref{e11}), we can derive the second-order equation for $\phi$.
Under this scheme of approximations, we obtain
\be
\frac{1}{r^2} \left( r^2 \phi' \right)'
\simeq \mu_1\rho_m+\mu_2\,, 
\label{fieldeq}
\ee
where 
\ba
\mu_1 &=&-\frac{\phi'r}{4A_4\beta}\left(A_{3,X}+\beta
+\frac{6A_{5,X}+B_{5,X}}{r^2} \right)\,,\label{mu1}\\
\mu_2 &=&\frac{1}{\beta r}
\Bigg[\left(\frac{A_{2,\phi}}{2}-X A_{2,\phi X}\right)r^2
+(A_{3,\phi}-2X A_{3,\phi X}+4\phi' XA_{2,XX})r\notag\\
&&+2\phi'(A_{3,X}+4X A_{3,XX})
-2A_{4,\phi}+2XA_{4,\phi X}
-\frac{8\phi' XA_{4,XX}+r \alpha_{1}
-4\phi' \alpha_{2}}{r} \Bigg]\,. \label{mu2}
\ea
The parameters $\beta$, $\alpha_1$, $\alpha_2$ 
are given by 
\ba
\beta&\equiv&
(A_{2,X}+2X A_{2,XX})r+2(A_{3,X}+2X A_{3,XX}) 
-\frac{4XA_{4,XX}}{r}+\frac{2\alpha_{2}}{r}\,,
\label{beta} \\
\alpha_1&\equiv&2XB_{4,\phi X}-B_{4,\phi}-A_{4,\phi}\,,\\
\alpha_2&\equiv&2XB_{4,XX}+B_{4,X}-A_{4,X}\,. 
\ea
{}From Eqs.~(\ref{e00}) and (\ref{e11}) it follows that 
\ba
\left( r\Phi \right)' &\simeq& -\frac{\rho_m r^2}{4A_4}+\frac{r^2}{4A_4}
\biggl[ A_2-\phi' \left( A_{3,\phi}+
2\phi'' A_{3,X} \right)
+\frac{4\phi'}{r} \left( A_{4,\phi}+
2\phi'' A_{4,X} \right) \nonumber \\
& &+\frac{\phi'}{r^2} \left\{ 2\phi'' 
\left( 6A_{5,X}+B_{5,X} \right)+
6A_{5,\phi}+B_{5,\phi} \right\}\biggr] 
-\frac12 \alpha_{{\rm t}4}\,,\label{gra1}\\
\Psi' &\simeq& \frac{\Phi}{r}-\frac{r}{4A_4} \left( A_2
-2{\phi'}^2 A_{2,X} \right)+\frac{{\phi'}^2  A_{3,X}}{A_4} 
-\frac{{\phi'}^2  A_{4,X}}{A_4r}
+\frac{\aH}{2r}\,.
\label{gra2}
\ea

In Eq.~(\ref{fieldeq}) the effect of the Lagrangian $L_5$ arises only
for the term $\mu_1 \rho_m$, which leads to an additional interaction
between the scalar field and matter.  For the theories in which the
condition
\be
6A_{5,X}+B_{5,X}=0
\label{ABcon}
\ee
is satisfied, Eq.~(\ref{fieldeq}) is of the same form as the one
without $L_5$.  If $A_5$ and $B_5$ depend on $X$ alone, it follows
that the term $(6A_{5,\phi}+B_{5,\phi})\phi'/r^2$ in Eq.~(\ref{gra1})
vanishes and that the $G_5$-dependent terms do not contribute to $A_4$
and $B_4$.  This means that, for $A_5(X)$ and $B_5(X)$ satisfying the
condition (\ref{ABcon}), the resulting profiles of $\phi,\Phi,\Psi$
are the same as those for the case $L_5=0$.

In Horndeski theories the relation $A_5=-XB_{5,X}/3$ holds, so the
condition (\ref{ABcon}) can be satisfied only for the specific
functions $A_5=cX^{1/2}$ and $B_5=-6cX^{1/2}$ ($c$ is a constant).
However, for the regular function $G_5$ given by Eq.~(\ref{G5ex}), we
have $A_5$ and $B_5$ in the forms (\ref{A5ex}) and (\ref{B5ex}) with
$F_5=0$, so the above specific case is not encompassed.  Hence we need
to go beyond the Horndeski domain to fulfill the condition
(\ref{ABcon}).

\subsection{Concrete model}

In the framework of GLPV theories, we consider a massless scalar field
$\phi$ coupled to the Ricci scalar $R$ with $F(\phi)R$.  We study how
non-linear field self-interactions like $f_4(X)$ in $G_4$ and $f_5(X)$
in $G_5$ affect the field profile inside/outside the body. We also
take into account the ``beyond-Horndeski'' terms $F_4=g_4(X)$ and
$F_5=g_5(X)$.  Namely, we study the theories given by
\ba
& &
G_2=-\frac12 X\,,\qquad G_3=0\,, \qquad
G_4=\frac12 M_{\rm pl}^2 F(\phi)+f_4(X)\,,\qquad
G_5=f_5(X)\,,\nonumber \\
& &
F_4=g_4(X)\,,\qquad F_5=g_5(X)\,.
\label{conmo}
\ea
On using the correspondence (\ref{A2})-(\ref{B5}) with $X>0$, these
functions translate to
\ba
\hspace{-0.5cm}
& &
A_2=-\frac12 X\,,\qquad A_3=M_{\rm pl}^2\sqrt{X} 
F_{,\phi}\,, \qquad
A_4=-\frac12 M_{\rm pl}^2 F-f_4+2Xf_{4,X}-X^2g_4\,,
\nonumber \\
\hspace{-0.5cm}
& &
B_4=\frac12 M_{\rm pl}^2 F+f_4\,,\qquad 
A_5=-X^{3/2} \left( Xg_5+\frac13 f_{5,X} \right)\,,
\qquad 
B_5=\int \sqrt{X} f_{5,X}dX\,.
\label{modelcon}
\ea
For concreteness, we focus on the following functions
\be
F=e^{-2q\phi/M_{\rm pl}}\,,\qquad
f_4=\frac{c_4}{M^6}X^2\,,\qquad
f_5=\frac{c_5}{M^9}X^2\,,\qquad
g_4=\frac{d_4}{M^6}\,,\qquad
g_5=\frac{d_5}{M^9}\,,
\label{Ffg}
\ee
where $M$ is a constant having a dimension of mass, and
$q,c_4,c_5,d_4,d_5$ are dimensionless constants whose magnitudes are
at most of the order of 1.  The parameter $\aH$ is given by
\be
\aH=-\frac{2d_4X^2}{M^6M_{\rm pl}^2F
-2(3c_4-d_4)X^2}\,,
\ee
which vanishes as $r \to 0$ for the boundary condition
$\phi'(0)=0$. Hence there is no conical singularity at $r=0$ induced
by $\aH$.  The functions $A_5$ and $B_5$ read
\be
A_5=-\frac{2c_5+3d_5}{3M^9}X^{5/2} \,,\qquad
B_5=\frac{4c_5}{5M^9}X^{5/2}\,,
\label{A5B5}
\ee
which obey the relation 
$A_5+XB_{5,X}/3=-d_5M^9X^{5/2}$. 
The functions (\ref{A5B5}) correspond to $n=2$ and 
$F_5={\rm constant}$ in Eqs.~(\ref{A5ex})-(\ref{B5ex}), 
so no additional curvature singularity is induced 
by the Lagrangian $L_5$.

We fix the mass scale $M$ to be
\be
M^3=M_{\rm pl}H_0^2\,,
\label{Mscale}
\ee
where $H_0$ is the today's Hubble parameter whose inverse is of the
order of $H_0^{-1}\simeq 10^{28}$ cm.  The choice (\ref{Mscale}) is
associated with the fact that we are interested in the case where the
field $\phi$ is responsible for the present cosmic acceleration like 
the covariant Galileon model \cite{covariant,DT10}. 
The Galileon cosmology studied in Ref.~\cite{DT10} is based on 
the minimal coupling model ($F(\phi)=1$) 
with $g_4=0=g_5$, in which case there is a tracker that finally
approaches a de Sitter attractor characterized 
by $X={\rm constant}$.
The model (\ref{conmo}) is the extension of covariant
Galileons to the case in which there are two more functions $g_4$ and
$g_5$.  The coupling between the field $\phi$ and gravity is
characterized by the non-zero constant $q$.  The field is indirectly
coupled to matter through its gravitational interaction.

For the above theories, the field equation of motion is 
given by Eq.~(\ref{fieldeq}) with 
\ba
\mu_1 &=&
-\frac{1}{2\beta r} \frac{(qM_{\rm pl}F-\beta \phi')r^2
+(8c_5+15d_5){\phi'}^4/M^9}
{M_{\rm pl}^2F-2(3c_4-d_4){\phi'}^4/M^6}\,,
\label{mu1con}\\
\mu_2 &=& 
-\frac{24(2c_4-d_4)}{\beta} \frac{{\phi'}^3}{M^6r^2}\,,\\
\beta &=& 
-\frac{r}{2} \left[ 1+\frac{24(2c_4-d_4){\phi'}^2}
{M^6r^2} \right]\,.
\label{betacon}
\ea

We derive the solution to Eq.~(\ref{fieldeq}) around a compact body
with the $r$-dependent density $\rho_m(r)$.  For concreteness, let us
consider the density distribution
\be
\rho_m(r)=\rho_c\,e^{-r^2/r_t^2}\,,
\label{profile}
\ee
where $\rho_c$ is the central density.  The density starts to vary
significantly at the transition radius $r_t$, which is of the same
order as the radius $r_s$ of the body.  We can also consider other
density profiles which decrease for larger $r$, but the qualitative
behavior of solutions is similar to that discussed below.

To derive the field profile analytically as well as numerically, it is
convenient to introduce the following dimensionless quantities
\ba
& &
x=\frac{r}{r_s}\,,\qquad
y=\frac{M_{\rm pl}{\phi'}^3(r)}
{M^6 \rho_c r_s^3}\,,\qquad
z=\frac{\phi}{M_{\rm pl}}\,, \nonumber \\
& &
\lambda_1=\left( \frac{\rho_c r_s^2}
{M_{\rm pl}^2} \right)^{1/3}\,,\qquad
\lambda_2=\left( \frac{M^3r_s^2}{M_{\rm pl}} 
\right)^{1/3}\,,\qquad 
\xi_t=\frac{r_t}{r_s}\,.
\ea
Then, Eq.~(\ref{fieldeq}) can be written as
\ba
\frac{dy}{dx}
&=&-\frac{6}{x}y
+\frac{144s_4 \lambda_1^2y^{5/3}}
{x(\lambda_2^2 x^2
+24s_4 \lambda_1^2 y^{2/3})} \nonumber \\
& &+\frac{3}{2}
\frac{\lambda_2(2qFx^2+\lambda_1\lambda_2^2 x^3y^{1/3}
+24\lambda_1^3 s_4xy)+2s_5\lambda_1^4y^{4/3}}
{\lambda_2(\lambda_2^2x^2+24s_4\lambda_1^2 y^{2/3})
[F-2\lambda_1^4\lambda_2^2 (3c_4-d_4)y^{4/3}]}
\lambda_1^2\,e^{-x^2/\xi_t^2}\,y^{2/3}\,,
\label{dyx}
\ea
where 
\be
s_4 \equiv 2c_4-d_4\,,\qquad
s_5 \equiv 8c_5+15d_5\,.
\ee
We remind the reader that $d_5=0$ corresponds to the case $F_5=0$ 
(one of the two conditions Horndeski theories satisfy), whereas
$s_5=0$ implies the case in which $L_5$ does not affect the
dominant term in the scalar-field equation (\ref{fieldeq}). 
The quantity $z$ obeys the differential equation
\be
\frac{dz}{dx}=\lambda_1 \lambda_2^2\,y^{1/3}\,.
\label{dzx}
\ee

Since $\lambda_2/\lambda_1=(M_{\rm pl}^2
H_0^2/\rho_c)^{1/3} \simeq (\rho_0/\rho_c)^{1/3}$, 
where $\rho_0=10^{-29}$ g/cm$^3$ is the today's
cosmological density, $\lambda_2$ is much smaller than 
$\lambda_1$ for compact bodies like the Sun 
or the Earth. For the Sun ($\rho_c \simeq 100$ g/cm$^3$ 
and $r_s \simeq 7 \times 10^{10}$ cm), we have 
that $\lambda_1 \simeq 0.1$ and 
$\lambda_2 \simeq 10^{-12}$. 
If we consider the distance $x=r/r_s$ 
for which the condition 
\be
\lambda_2^2 x^2 \ll | 24s_4 \lambda_1^2 
y^{2/3}|
\label{lamcon}
\ee
is satisfied, the first two terms on the r.h.s. of 
Eq.~(\ref{dyx}) cancel each other. 
In this regime only 
the density-dependent term survives, so that 
Eq.~(\ref{dyx}) is simplified to
\be
\frac{dy}{dx} \simeq 
\frac{\lambda_2(2qFx^2+\lambda_1\lambda_2^2 x^3y^{1/3}
+24\lambda_1^3 s_4xy)+2s_5\lambda_1^4y^{4/3}}{16s_4 \lambda_2
[F-2\lambda_1^4\lambda_2^2 (3c_4-d_4)y^{4/3}]} 
e^{-x^2/\xi_t^2}\,.
\label{dyx2}
\ee
The distance at which the condition (\ref{lamcon}) starts to be
violated corresponds to the Vainshtein radius $r_V$.  This radius can
be derived after obtaining the solution to Eq.~(\ref{dyx2}).  Provided
the Vainshtein mechanism is at work, the variation of
$z=\phi/M_{\rm pl}$ is tiny, such that the quantity $F=e^{-2qz}$ stays
nearly constant.  We shall consider the boundary condition of $z(r=0)$
where $F$ is close to 1.

The condition (\ref{ABcon}) translates to $8c_5+15d_5=0$, under which
the contributions from $L_5$ to Eqs.~(\ref{fieldeq}) and (\ref{gra1})
vanish.  In Sec.~\ref{s50sec} we first discuss the solution to the
field in this case and proceed to the theories with
$d_5 \neq -8c_5/15$ in Sec.~\ref{s5neq0}.
 
Before deriving the field solution for each case, we provide the
general argument about the behavior of gravitational potentials for
the model given by the functions (\ref{modelcon}).  Suppose that the
solution to the field $\phi$ around the compact body is expressed in
the form
\be
\phi'(r)=m^2 \left( \frac{r}{r_s} \right)^p\,,
\label{phias}
\ee
where $m$ corresponds to some mass scale and $p~(\geq 0)$ is a
constant. Then, we can integrate Eqs.~(\ref{gra1})-(\ref{gra2}) to
obtain $\Phi$ and $\Psi$.  In doing so, we define the Schwarzschild
radius of the source, as
\be
r_g=\frac{1}{M_{\rm pl}^2}\int_0^{r} 
\rho_m (\tilde{r}) \tilde{r}^2 d\tilde{r}\,.
\ee

For $p >0$, substitution of Eq.~(\ref{phias}) into
Eqs.~(\ref{gra1})-(\ref{gra2}) gives
\ba
\Phi(r) &\simeq&
\frac{r_g}{2r} \biggl[ 1-2q\frac{(mr)^2}{M_{\rm pl}r_g}x^p
+\frac{(1+8q^2)m^4r^3}{2(2p+3)M_{\rm pl}^2r_g}x^{2p} 
\nonumber \\
& &~~~~~
-\frac{2\{ 4c_4(1+6p)-d_4(1+8p) \}m^8r}
{(4p+1)M^6M_{\rm pl}^2r_g}x^{4p} +\frac{2s_5m^{10}}
{5M^9M_{\rm pl}^2r_g}x^{5p}
\biggr]\,,\label{Phige}\\
\Psi(r) &\simeq&
-\frac{r_g}{2r} \biggl[ 1-2q\frac{(mr)^2}
{(1+p)M_{\rm pl}r_g}x^p
-\frac{(2+p+4q^2)m^4r^3}
{2(3+5p+2p^2)M_{\rm pl}^2r_g}x^{2p} 
\nonumber \\
& &~~~~~~
-\frac{2s_4(3p+1)m^8r}
{p(4p+1)M^6M_{\rm pl}^2r_g}x^{4p} 
-\frac{2s_5m^{10}}
{5(5p-1)M^9M_{\rm pl}^2r_g}x^{5p}
\biggr]\,,\label{Psige}
\ea
where the integration constant arising in Eq.~(\ref{Phige}) has been
absorbed into $r_g$, and another integration constant arising in
Eq.~(\ref{Psige}) has been set to 0 to satisfy the boundary condition
$\Psi(r \to \infty)=0$. We have also neglected the term
$(3c_4-d_4)X^2/M^6$ in $A_4$, which gives rise to the contributions to
$\Phi$ and $\Psi$ of the orders of
$(3c_4-d_4)m^8x^{4p}/(M^6M_{\rm pl}^2)r_g/r$ (which are smaller than
the fourth terms in Eqs.~(\ref{Phige}) and (\ref{Psige}) for $r>r_g$).

For $p=0$, the solutions to $\Phi$ and $\Psi$
are given by 
\ba
\Phi(r) &\simeq&
\frac{r_g}{2r} \left[ 1-2q\frac{(mr)^2}{M_{\rm pl}r_g}
+\frac{(1+8q^2)m^4r^3}{6M_{\rm pl}^2r_g}
-\frac{2(4c_4-d_4)m^8r}{M^6M_{\rm pl}^2 r_g}
\right]\,,\label{Phige2}\\
\Psi(r) &\simeq&
-\frac{r_g}{2r} \left[ 1-2q\frac{(mr)^2}
{M_{\rm pl}r_g}
-\frac{(1+2q^2)m^4r^3}{3M_{\rm pl}^2r_g}
-\frac{8(2c_4-d_4)m^8}{M^6M_{\rm pl}^2} 
\frac{r}{r_g} \ln \frac{r}{r_*}
\right]\,,\label{Psige2}
\ea
where $r_*$ is an integration constant.  Since the $\phi''$ term in
Eq.~(\ref{gra1}) vanishes in this case, the Lagrangian $L_5$ does not
affect the gravitational potentials for the constant $\phi'$ solution.

For the consistency with local gravity experiments, the post Newtonian
parameter $\gamma=-\Phi/\Psi$ is constrained to be \cite{Will}
\be
\left| \gamma-1 \right|<2.3 \times 10^{-5}
\label{solar}
\ee
inside the solar system.

\subsection{Model with $s_5=0$}
\label{s50sec}

Let us first study the field profile and the gravitational potentials
for the model with $s_5=0$, i.e., $8c_5+15d_5=0$.  In this case, the
$L_5$-dependent term vanishes from the field equation. Around the
compact body (including both interior and exterior), we employ the
approximations that
$F \gg |2\lambda_1^4\lambda_2^2 (3c_4-d_4)y^{4/3}|$ and that the
second term in the numerator of Eq.~(\ref{dyx2}) is sub-dominant
relative to the first term.  The consistency of these approximations
can be checked after deriving the solution.  Then, Eq.~(\ref{dyx2}) is
further simplified to
\be
\frac{dy}{dx} \simeq 
\frac{1}{8s_4F} \left( qFx^2
+12 \lambda_1^3 s_4xy \right)e^{-x^2/\xi_t^2}\,.
\label{dyx3}
\ee
Under the boundary condition $\phi'(0)=0$, 
the solution to Eq.~(\ref{dyx2}) around 
$r=0$ (at which $e^{-x^2/\xi_t^2} \simeq 1$)
is given by $y(x)=qx^3/(24s_4)$, i.e., 
\be
\phi'(r)= \left( \frac{q\rho_cM^6}
{24s_4 M_{\rm pl}} \right)^{1/3}r\,,
\qquad \quad (0 < r \ll r_s).
\label{phir=0}
\ee

On using the solution (\ref{phir=0}), 
we find that the second term on the r.h.s. 
of Eq.~(\ref{dyx3}) is smaller than 
the first one for the distance 
\be
r< r_s \sqrt{\frac{2F}
{\lambda_1^3}} \equiv r_1\,.
\label{r1con}
\ee
For $\lambda_1 \lesssim O(1)$, which includes the case of the Sun,
$r_1$ is larger than $r_s$.  Around the surface of the body, however,
the exponential term $e^{-x^2/\xi_t^2}$ starts to suppress the
r.h.s. of Eq.~(\ref{dyx3}).  Employing the approximation that the
r.h.s. of Eq.~(\ref{dyx3}) vanishes in the regime $r>r_s$, we obtain
$\phi'(r)={\rm constant}$.  Matching this with Eq.~(\ref{phir=0}) at
$r=r_s$, the solution for $r>r_s$ is given by
\be
\phi'(r)= \left( \frac{q\rho_cM^6}
{24s_4 M_{\rm pl}} \right)^{1/3}r_s\,,
\qquad \quad (r_s < r <r_V).
\label{phir1}
\ee

The solution (\ref{phir1}) is valid up to the 
radius $r_V$ at which the condition (\ref{lamcon}) 
starts to be violated. This radius can be estimated as
\be
r_V= \left( 24|s_4| \right)^{1/6} 
\left( \frac{|q|\rho_c}{M^3 M_{\rm pl}} 
\right)^{1/3}r_s\,.
\label{rVes}
\ee
In the regime $r>r_V$, the first term $-6y/x$ on the r.h.s. of 
Eq.~(\ref{dyx}) dominates over the other terms, so 
we obtain the solution $\phi'(r) \propto 1/r^2$. 
Matching this with Eq.~(\ref{phir1}), 
the resulting solution for $r>r_V$ is given by 
\be
\phi'(r)= \left( \frac{q\rho_cM^6}
{24s_4 M_{\rm pl}} \right)^{1/3}
\frac{r_s r_V^2}{r^2}\,,
\qquad \quad (r >r_V).
\label{phir2}
\ee
The Schwarzschild radius for the density profile 
(\ref{profile}), in the limit $r \to \infty$, reads
\be
r_g=\frac{\sqrt{\pi}}{4} \frac{\rho_cr_t^3}
{M_{\rm pl}^2} \approx 
\frac{\rho_cr_s^3}
{M_{\rm pl}^2}\,,
\ee
where we used the approximation $r_t \approx r_s$. 
Then, the solutions (\ref{phir1}) and (\ref{phir2}) 
can be expressed in the forms 
$\phi'(r) \approx qM_{\rm pl}r_g/r_V^2$ 
and $\phi'(r) \approx qM_{\rm pl}r_g/r^2$, respectively,  
with $r_V \approx |s_4|^{1/6} (|q|M_{\rm pl}r_g)^{1/3}/M$.
These solutions coincide with those for the theories 
with $L_5=0$ \cite{AKT15}.

\begin{figure}
\begin{center}
\includegraphics[width=4.0in]{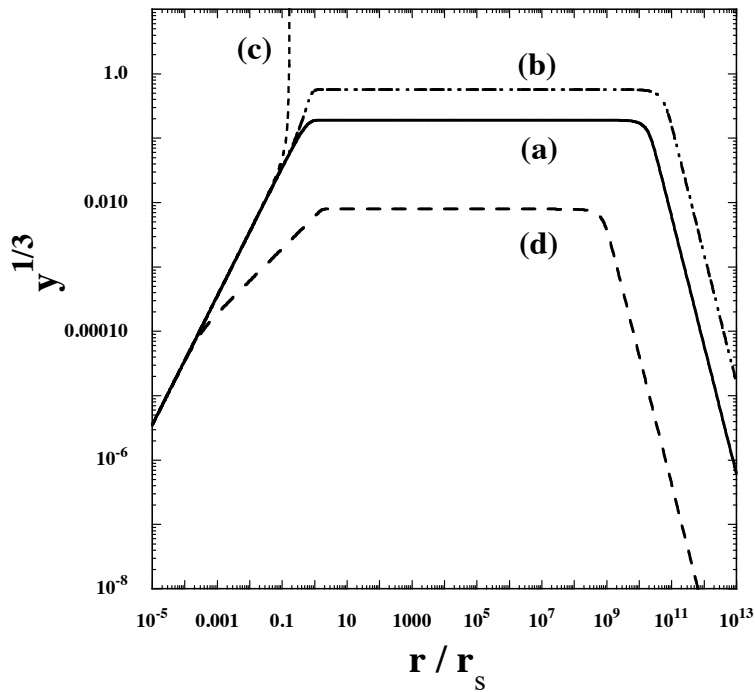}
\caption{The field profile $y^{1/3}=
[M_{\rm pl}/(M^6\rho_cr_s^3)]^{1/3}\phi'(r)$ 
versus $r/r_s$ for $q=1$, $\lambda_1=0.1$, 
$\lambda_2=3.0 \times 10^{-12}$, $\xi_t=0.5$, 
$c_4=1$, and $d_4=1$. 
Each case corresponds to 
(a) $c_5=1$, $d_5=-8/15$, 
(b) $c_5=1.0 \times 10^{-6}$, 
$d_5=1.0 \times 10^{-6}$, 
(c) $c_5=1.0 \times 10^{-5}$, 
$d_5=1.0 \times 10^{-5}$,
(d) $c_5=-1.0$, $d_5=-1.0$. 
We have chosen initial conditions $y=10^{-40}$ and $z=0$ 
at $r/r_s=e^{-15}$.
The Vainshtein mechanism is most effective for the case (d). 
\label{fig1}}
\end{center}
\end{figure}

In Fig.~\ref{fig1} we plot an example of the field profile for $c_5=1$
and $d_5=-8/15$ (i.e., $s_5=0$) as the case (a) with the model
parameters $c_4=1$, $d_4=1$, $q=1$, $\lambda_1=0.1$,
$\lambda_2=3.0 \times 10^{-12}$, $\xi_t=0.5$ (the density profile of
the Sun is modeled).  This profile is derived by numerically
integrating Eqs.~(\ref{dyx}) and (\ref{dzx}) with the boundary conditions
$y(r)=10^{-40}\approx0$ and $z(r)=0$ at $r/r_s=e^{-15}\approx0$. 
The behavior of the solutions does not strongly depend on 
the choice of the boundary conditions. 
As estimated by Eqs.~(\ref{phir=0}), (\ref{phir1}), and (\ref{phir2}),
the field derivative $\phi'(r)$ linearly grows for $0<r \lesssim r_s$,
stays with a constant value for $r_s \lesssim r<r_V$, and then
decreases as $\propto 1/r^{2}$ for $r>r_V$.  For the Sun
($r_s=7 \times 10^{10}$ cm) with $|s_4|=O(1)$, the Vainshtein radius
can be estimated as $r_V \simeq 3 \times 10^{20}$ cm.

For the exterior solution (\ref{phir1}), the mass scale $m$ in
Eq.~(\ref{phias}) reads
\be
m^2=\left( \frac{q\rho_cM^6}
{24s_4 M_{\rm pl}} \right)^{1/3}r_s
\approx \frac{qM_{\rm pl}r_g}{r_V^2}\,,
\label{ma1}
\ee
with the power $p=0$.  Substituting Eq.~(\ref{ma1}) into
Eqs.~(\ref{Phige2})-(\ref{Psige2}), we obtain the gravitational
potentials in the regime $r_s<r<r_V$.  For the Sun
($r_g\simeq 3 \times 10^5$ cm) with $|q|, |c_4|, |d_4|$ of the order
of unity, the second, third, fourth terms on the r.h.s.  of
Eqs.~(\ref{Phige2}) and (\ref{Psige2}) do not exceed the orders of
$10^{-13}$, $10^{-34}$, and $10^{-20}$, respectively, inside the solar
system ($r \lesssim 10^{14}$~cm).  Hence the solar-system constraint
(\ref{solar}) is satisfied for the field solution in the regime
$r_s<r<r_V$.
 
The interior solution (\ref{phir=0}) corresponds to the mass $m$ same
as Eq.~(\ref{ma1}) with $p=1$, in which case the gravitational
potentials are given by Eqs.~(\ref{Phige})-(\ref{Psige}) with $s_5=0$.
The second, third, fourth terms on the r.h.s.  of Eqs.~(\ref{Phige})
and (\ref{Psige}) are at most of the orders of $10^{-19}$, $10^{-44}$,
and $10^{-24}$, respectively, around the surface of the Sun, so the
bound (\ref{solar}) is well satisfied.

%
\subsection{Model with $s_5 \neq 0$}
\label{s5neq0}

Let us proceed to the discussion of solutions for non-zero values of
$s_5$.  In high-density regions with the distance satisfying
the condition (\ref{lamcon}), we can employ the same approximations as
those used for the derivation of Eq.~(\ref{dyx3}). For the model with
$s_5 \neq 0$, Eq.~(\ref{dyx3}) is modified to
\be
\frac{dy}{dx} \simeq 
\frac{1}{8\lambda_2 s_4F} \left( 
qF\lambda_2 x^2+12 \lambda_1^3 \lambda_2 
s_4 xy+\lambda_1^4 s_5 y^{4/3}
\right)e^{-x^2/\xi_t^2}\,.
\label{dyx4}
\ee
Around the center of the body ($0<r \ll r_s$) the first term on the
r.h.s. of Eq.~(\ref{dyx4}) is the dominant contribution, so we obtain
the same solution as Eq.~(\ref{phir=0}), i.e.,
\be
y(x)=\frac{q}{24s_4}x^3\,.
\label{ysmall}
\ee
We recall that the second term is sub-dominant to the first one under
the condition (\ref{r1con}).  The third term is smaller than the first
one for the distance
\be
r<\sqrt{\frac{F\lambda_2}{|s_5|\lambda_1^4}} 
\frac{(24|s_4|)^{2/3}}{|q|^{1/6}}
r_s \equiv r_2\,.
\label{r2con}
\ee
Unless $|s_5|$ is much less than unity, we have that
$r_2 \ll \{ r_s, r_1 \}$ for $\lambda_2 \ll 1$ and for
$\lambda_1, |s_4|, |q|$ of the order of unity.  Then, the
$L_5$-dependent contribution begins to be important for the distance
$r_2$ much smaller than $r_1$. In what follows, we shall discuss the
two cases (a) $s_5>0$ and (b) $s_5<0$ separately, for positive values
of $q$ and $s_4$.

\subsubsection{$s_5>0$}

If $s_5>0$, then the third term on the r.h.s.\ of Eq.~(\ref{dyx4}) 
dominates over the first one for the distance $r>r_2$. 
Picking up this dominant contribution alone, 
we obtain the following integrated solution 
\be
y(x)^{-1/3}+\frac{\sqrt{\pi}\lambda_1^4s_5}
{48\lambda_2 s_4F}\,\xi_t\, 
{\rm erf} \left( \frac{x}{\xi_t} \right)=C\,,
\label{yxin}
\ee
where ${\rm erf}(x)=(2/\sqrt{\pi})\int_0^x e^{-s^2}ds$ 
is the error function and 
$C$ is an integration constant. 
Since the solution (\ref{ysmall}) is valid up to 
the dimensionless distance $x_2 \equiv r_2/r_s$, 
we can match it with Eq.~(\ref{yxin}) at $x=x_2$.
Then, the solution for the distance $r>r_2$ (but for $r$ 
smaller than the Vainshtein radius $r_V$ derived later) 
is given by 
\be
y(x)=\left[ \left( \frac{24s_4}{qx_2^3} \right)^{1/3} 
+\frac{\sqrt{\pi}\lambda_1^4s_5}
{48\lambda_2 s_4F}  \xi_t  
\left\{ {\rm erf}  \left( \frac{x_2}{\xi_t} \right)
-{\rm erf}  \left( \frac{x}{\xi_t} \right) \right\}
\right]^{-3}\,.
\label{yx1}
\ee
Provided that $r_2 \ll r_t$, the term
$\xi_t\,{\rm erf} (x_2/\xi_t)$ is close to 0.
The function $y(x)$ increases for larger $x$
with the growth of the error function 
${\rm erf} (x/\xi_t)$ toward 1.

If the condition 
\be
s_5<
\left( \frac{24s_4}{qx_2^3} \right)^{1/3} 
\frac{48\lambda_2s_4F}{\sqrt{\pi}\lambda_1^4\xi_t}
\label{s5con}
\ee
is satisfied, then $y(x)$ approaches the value
\be
y(x) \to \left[ \left( \frac{24s_4}{qx_2^3} \right)^{1/3} 
-\frac{\sqrt{\pi}\lambda_1^4\xi_ts_5}
{48\lambda_2 s_4F} 
\right]^{-3}\,,
\label{yx2}
\ee
where we used the approximation
$\xi_t {\rm erf} (x_2/\xi_t) \simeq 0$.  On using Eq.~(\ref{r2con}),
one can eliminate $x_2$ in Eq.~(\ref{s5con}).  Then, the condition
(\ref{s5con}) translates to
\be
s_5<s_5^{\rm max} \equiv \frac{192}{\pi} \left( \frac{3s_4^4}{q} 
\right)^{1/3}\frac{\lambda_2}{\lambda_1^4} 
\frac{F}{\xi_t^2}\,.
\label{s5con2}
\ee
{}From Eq.~(\ref{yx2}) it follows that the positive value of $s_5$
leads to the enhancement of $\phi'(r)$.  The case (b) shown in
Fig.~\ref{fig1} corresponds to such an example, in which case the
transition to the solution (\ref{yx2}) occurs around
$r_2 \approx r_s$.  We caution that the solution (\ref{yx2}) is valid
for the distance $r$ satisfying the condition (\ref{lamcon}), i.e.,
\be
r<r_V \equiv (24 s_4)^{1/6} \left( 
\frac{q \rho_c}{M^3M_{\rm pl}} \right)^{1/3}
\left( 1-\frac{s_5}{s_5^{\rm max}} 
\right)^{-1} r_2\,.
\label{rVes2}
\ee
Compared to the Vainshtein radius (\ref{rVes}) derived for $s_5=0$,
the existence of the factor $(1-s_5/s_5^{\rm max})^{-1}$ in
Eq.~(\ref{rVes2}) works to increase $r_V$ for $s_5>0$.  In fact, this
property can be seen by comparing the cases (a) and (b) in
Fig.~\ref{fig1}.  Outside the Vainshtein radius the $s_5$-dependent
term in Eq.~(\ref{dyx}) is suppressed, so integration of the equation
$dy/dx \simeq -6y/x$ gives the solution $y(x) \propto 1/x^6$.
Matching this with Eq.~(\ref{yx2}), we obtain the solution for
$r>r_V$:
\be
y(x)=\frac{qx_2^3}{24s_4} 
\left(1-\frac{s_5}{s_5^{\rm max}}\right)^{-3} 
\left(\frac{r_V}{r_s}\right)^6 \frac{1}{x^6}\,. 
\label{yx3}
\ee
In summary, provided the parameter $s_5$ is in the range
(\ref{s5con2}), the solution to $y(x)$ is given by Eq.~(\ref{ysmall})
for $0<r<r_2$, Eq.~(\ref{yx1}) for $r_2<r<r_V$, and Eq.~(\ref{yx3})
for $r>r_V$.

Unless $s_5$ is very close to $s_5^{\rm max}$, the order of $\phi'(r)$
is not very much different from that for $s_5=0$, see the cases (a)
and (b) in Fig.~\ref{fig1}.  Hence the screening of the fifth force in
the regime $r_s<r<r_V$ works in a similar way to that discussed in
Sec.~\ref{s50sec}.  Provided that $y(x)$ given by Eq.~(\ref{yx2}) is
not much different from the value $qx_2^3/(24s_2)$, the solar-system
constraint (\ref{solar}) is satisfied for $r_s<r<r_V$.

For the distance $r<r_s$ the variation of $\phi'(r)$ occurs, so the
Lagrangian $L_5$ gives rise to non-vanishing contributions to the
gravitational potentials.  In the case of the Sun, the last terms of
Eqs.~(\ref{Phige}) and (\ref{Psige}) are at most of the order of
$10^{-15}s_5$.  Meanwhile, the condition (\ref{s5con2}) translates to
$s_5 \lesssim 3 \times 10^{-6}\,s_4^{4/3}F/(q^{1/3}\xi_t^2)$.  
For $s_4$, $q$, $\xi_t$ of the order of unity, the corrections to 
$\Phi$ and $\Psi$ induced by the non-zero $s_5$ term are 
smaller than $10^{-21}$.
 
If the condition (\ref{s5con2}) is violated, the field derivative
exhibits the divergence at the distance $r_{\rm D}$ satisfying
\be
{\rm erf} \left( \frac{r_{\rm D}}{r_t} \right)=
\frac{s_5^{\rm max}}{s_5}\,.
\ee
In this case, the screening mechanism is disrupted by the presence of
the Lagrangian $L_5$.  In the case (c) of Fig.~\ref{fig1}, the field
derivative goes to infinity at the distance $r_{\rm D}$ smaller than
$r_s$. For larger $s_5$, $r_{\rm D}$ gets smaller.  To avoid the
divergent behavior of $\phi'(r)$, we require that the parameter $s_5$
is in the range (\ref{s5con2}).  In the numerical simulation of
Fig.~\ref{fig1} we have chosen the values $s_4=1$, $q=1$, and
$\xi_t=0.5$, so the condition (\ref{s5con2}) translates to
$s_5<{\cal O}(10^{-5})$ for the Sun.

\subsubsection{$s_5<0$}

Let us proceed to the discussion for the case $s_5<0$. 
The third term on the r.h.s. of Eq.~(\ref{dyx4}), 
which is negative, tends to catch up with the first one 
around the distance $r=r_2$, but the solution is
saturated around the point $dy/dx=0$.  In the regime
$r_2<r< r_t \approx r_s$ at which the factor 
$e^{-x^2/\xi_t^2}$ is of the order of 1, 
the leading-order solution to Eq.~(\ref{dyx4}) is given by
\be
y(x)= \left( 
\frac{qF\lambda_2}{\lambda_1^4|s_5|} 
\right)^{3/4} x^{3/2}\,.
\label{yxn1}
\ee
One can take into account the next-to-leading order correction
$\epsilon(x)$ by expressing the solution in the form
$y(x)=y_0(x) [1-\epsilon(x)]$, 
where $y_0(x)$ is the leading-order
term (\ref{yxn1}).  Integrating Eq.~(\ref{dyx4}) under the condition
$|\epsilon(x)| \ll 1$, we obtain
$\epsilon(x)=[1-\mu {\cal C}\exp(-\mu x^{3/2})]/(\mu x^{3/2})$, where
$\mu=[q\lambda_1^{12}|s_5|^3/(\lambda_2^3s_4^4F^3)]^{1/4}/9$ and
${\cal C}$ is an integration constant.  The correction $\epsilon(x)$
decreases for increasing $x$, so $y(x)$ approaches the leading-order
solution (\ref{yxn1}).

The solution (\ref{yxn1}) is valid only in the case where $r_2$ given
by Eq.~(\ref{r2con}) is smaller than $r_s$. This translates to the
following condition
\be
\left| s_5 \right|>\frac{F\lambda_2}{\lambda_1^4}
\frac{(24|s_4|)^{4/3}}{|q|^{1/3}}\,.
\label{s5value}
\ee
For the Sun with $q,s_4$ of the order of 1, this condition
corresponds to $|s_5|>{\cal O}(10^{-6})$.  
If $|s_5|$ is smaller than
the r.h.s. of Eq.~(\ref{s5value}), then the solution (\ref{ysmall})
directly connects to the constant $y(x)$ solution around $r=r_s$.  In
this latter case ($|s_5|<{\cal O}(10^{-6})$ for the Sun), the field
profile is similar to that for $s_5=0$.

In what follows, we derive the field solution under the condition
(\ref{s5value}).  For $r \gtrsim r_s$, the exponential term
$e^{-x^2/\xi_t^2}$ leads to the suppression of the r.h.s. of
Eq.~(\ref{dyx4}), so $y(x)$ approaches a constant. Matching this
solution with Eq.~(\ref{yxn1}) at $r=r_s$, we obtain the following
solution for the distance $r_s \lesssim r<r_V$ :
\be
y(x)=\left( 
\frac{qF\lambda_2}{\lambda_1^4|s_5|} 
\right)^{3/4} \,.
\label{yxn2}
\ee
The Vainshtein radius $r_V$ corresponds to the distance at which the
condition (\ref{lamcon}) starts to be violated, i.e.,
\be
r_V=2\sqrt{6} \left( \frac{qF s_4^2}{\lambda_2^3|s_5|} 
\right)^{1/4}r_s\,.
\ee
Matching Eq.~(\ref{yxn2}) with the solution $y(x) \propto 1/x^6$ at
$r=r_V$, the solution for $r>r_V$ reads
\be
y(x) =\left( 
\frac{qF\lambda_2}{\lambda_1^4|s_5|} 
\right)^{3/4} \left(\frac{r_V}{r_s}\right)^6\frac{1}{x^6} \,.
\label{yxn3}
\ee

We recall that, for $s_5=0$, the field solution is given by
$y(x) \simeq q/(24s_4)$ for the distance $r_s<r<r_V$.  On using the
solution (\ref{yxn2}), the ratio between $\phi'(r)$ for the $s_5<0$
case and the $s_5=0$ case is given by
\be
\frac{\phi'(r)|_{s_5<0}}{\phi'(r)|_{s_5=0}}
=24^{1/3} \left( \frac{F^3}{q} \right)^{1/12}
\frac{\lambda_2^{1/4}}{\lambda_1}
\frac{s_4^{1/3}}{|s_5|^{1/4}}\,,
\label{phiratio}
\ee
in the regime $r_s \lesssim r<r_V$.  For the Sun with $q,s_4$ of the
order of 1, the r.h.s.  of Eq.~(\ref{phiratio}) can be estimated as
${\cal O}(10^{-2})|s_5|^{-1/4}$.  Under the condition (\ref{s5value}),
i.e., $|s_5|>{\cal O}(10^{-6})$, $\phi'(r)|_{s_5<0}$ is smaller than
$\phi'(r)|_{s_5=0}$.  For increasing $|s_5|$, both $\phi'(r)|_{s_5<0}$
and $r_V$ decrease in proportion to $|s_5|^{-1/4}$.

The case (d) shown in Fig.~\ref{fig1} corresponds to the field profile
for $c_5=d_5=-1.0$.  As estimated above, the field derivative in the
regime $r_s \lesssim r<r_V$ is suppressed relative to the $s_5=0$
case.  In the case (d) of Fig.~\ref{fig1}, the transition from the
solution (\ref{ysmall}) to another solution (\ref{yxn1}) occurs at the
distance $r_2$ much smaller than $r_s$.  In the regime
$r_2<r \lesssim r_s$ the field derivative increases as
$\phi'(r) \propto r^{1/2}$, whose growth rate is smaller than that for
$s_5=0$ ($\phi'(r) \propto r$).  This is the reason why the negative
values of $s_5$ lead to the suppression of $\phi'(r)$ for the distance
$r$ larger than $r_2$.

Since the constant solution of $\phi'(r)$ in the range
$r_s \lesssim r<r_V$ is smaller than that for $s_5=0$, the model with
$s_5<0$ is consistent with local gravity constraints.  For the
distance $r_2<r \lesssim r_s$, the solutions to gravitational
potentials are given by Eqs.~(\ref{Phige}) and (\ref{Psige}) with
$m^2=(qF\lambda_2/|s_5|)^{1/4}\lambda_2^2 M_{\rm pl}/r_s$ and
$p=1/2$. For the Sun with $q,c_4,d_4$ of the order of 1, the second,
third, fourth, and fifth terms on the r.h.s. of Eqs.~(\ref{Phige}) and
(\ref{Psige}) can be estimated as $10^{-20}|s_5|^{-1/4}$,
$10^{-47}|s_5|^{-1/2}$, $10^{-29}|s_5|^{-1}$, and
$10^{-22}|s_5|^{-1/4}$, respectively, around $r=r_s$.  Under the
condition (\ref{s5value}), i.e., $|s_5|>{\cal O}(10^{-6})$, these
terms are significantly smaller than 1.  If $|s_5|<{\cal O}(10^{-6})$,
the solution to gravitational potentials is similar to that for
$s_5=0$.  Thus, the model with $s_5<0$ is compatible with the
solar-system bound (\ref{solar}).

\section{Conclusions}
\label{consec} 

We have investigated the problems of the conical singularity and the
screening mechanism in full GLPV theories.  If the deviation parameter
$\aH$ from Horndeski theories approaches a non-zero constant at the
center of a compact body (in the presence of a spherically symmetric
background and in the absence of a shift-symmetry for the scalar
field $\phi$), the conical singularity arises at $r=0$ due to a geometric
modification of the four-dimensional Ricci scalar $R$.  Using the
general expansions of $\Phi$, $\Psi$, and $\phi$ around $r=0$, which
are valid for compact objects with a regular boundary condition
$\phi'(r=0)=0$, we have shown that curvature scalar quantities
(including $R$) remain finite if and only if the 0-th order term
$\Phi_0$ and the 1-st order terms $\Phi_1, \Psi_1$ vanish
identically. The appearance of the conical singularity is associated
with the non-vanishing value of $\Phi_0$ induced by $\aH$.

We have studied the effect of the Lagrangian $L_5$ on the problem 
of the conical singularity. The function $G_5$ given by 
Eq.~(\ref{G5ex}) not only accommodates most of the theories 
proposed in the literature, but it also corresponds to the expansion 
of the regular function $G_5$ around $r=0$. 
In this case, the functions $A_5$ and $B_5$ 
are of the forms (\ref{A5ex}) and (\ref{B5ex}), respectively. 
We showed that the Lagrangian $L_5$ inside the Horndeski 
domain ($F_5=0$) does not affect the conical singularity 
induced by non-zero $\aH$. 

In GLPV theories, there is a freedom for choosing the more general
regular function $A_5$ in the form (\ref{A5sum}).  Then, the solutions
to gravitational potentials around $r=0$ can be expressed as
Eqs.~(\ref{PhiGL})-(\ref{PsiGL}) up to linear order in $r$.  Even if
it is possible to fix $\Phi_0=0$ by the boundary condition for the
function (\ref{A5sum}) with $\aH \neq 0$, there are non-vanishing
contributions from $\aH$ to the gravitational potentials (to $\Phi_1$
and $\Psi_1$) which lead to the divergence of curvature quantities at
$r=0$.  Thus, the curvature singularity originating from the non-zero
$\aH$ cannot be removed by the GLPV Lagrangian $L_5$.  Conversely, the
GLPV Lagrangian $L_5$ alone does not give rise to any additional
curvature singularity. In other words, the term $L_4$ is the only one
responsible for the presence of the curvature singularity in
the case $\aH (r \to 0) \neq0$. 

In the presence of a non-minimal coupling between $\phi$ and $R$, we
also derived the field profile and the gravitational potentials inside
and outside the body under the approximation of weak gravity. We
focused on the theories given by the functions (\ref{conmo}), in which
case no curvature singularities arise at $r=0$.  In GLPV theories
there is a specific model characterized by $s_5=8c_5+15d_5=0$, where
the contribution from the Lagrangian $L_5$ to the field equation
vanishes identically.  In this case the solutions to the gravitational
potentials are similar to those for the theories with $L_5=0$, so that
the model can be consistent with solar-system constraints under the
operation of the Vainshtein mechanism.

If the parameter $s_5$ is positive and the condition (\ref{s5con2}) is
violated, the field derivative $\phi'(r)$ exhibits divergent behavior
in high-density regions. As long as $s_5$ is smaller than the
order of $s_5^{\rm max}$, the screening mechanism can be at work
inside and outside the body.  If $s_5<0$ (which also includes, in
particular, the Horndeski case, $d_5=0$ with $c_5<0$),
then the field derivative is suppressed relative to the $s_5=0$ case
due to the existence of the intermediate solution (\ref{yxn1}) inside
the body (see the case (d) of Fig.~\ref{fig1}).  In this case, the
model is compatible with the solar-system bound (\ref{solar}).

We have thus shown that there are models in which local gravity
constraints can be satisfied in the framework of full GLPV theories
involving the Lagrangian $L_5$, while the curvature singularity is
absent at $r=0$.  It will be of interest to study whether or not such
models are cosmologically viable in connection to the dark energy
problem. We leave this for a future work.

\acknowledgments

RK is supported by the Grant-in-Aid for Research Activity Start-up 
of the JSPS No.\,15H06635. 
ST is supported by the Grant-in-Aid for Scientific Research Fund 
of the JSPS No.\,24540286, MEXT KAKENHI Grant-in-Aid for Scientific Research 
on Innovative Areas ``Cosmic Acceleration'' (No.\,15H05890),  
and the cooperation program between Tokyo
University of Science and CSIC.


\end{document}